\documentclass[aps,twocolumn,longbibliography]{revtex4-1}
\usepackage[colorlinks=true,citecolor=blue,linkcolor=blue,breaklinks=true]{hyperref}
\usepackage{epstopdf}
\usepackage{color}
\usepackage{bm}
\usepackage{graphicx}
\usepackage{amsmath}
\usepackage{amsfonts}
\usepackage{amssymb}
\usepackage{bezier} 
\usepackage{color} 
\usepackage{graphicx,amsmath,mathrsfs}
\usepackage{amssymb}
\usepackage{amsthm,multirow}
\usepackage{bm,bbm}
\usepackage{subfigure,xcolor} 
\usepackage[makeroom]{cancel}
\usepackage{color}
\usepackage{float}
\usepackage{bbold}
\def\bB{{\bf B}}

\def\bx{{\bf x}}

\def\br{{\bf r}}
\def\bz{{\bf z}}
\def\bj{{\bf j}}

\def\bA{{\bf A}}

\def\cL{{\cal L}}

\def\cO{{\cal O}}

\def\cM{{\cal M}}

\def\bP{{\bf P}}

\def\Im{{\mbox{Im}}}
\def\Re{{\mbox{Re}}}
\def\Tr{{\mbox{Tr}}}

\def\ij{{\langle ij\rangle}}

\def\be{\begin{equation}}
\def\ee{\end{equation}}
\def\bea{\begin{eqnarray}}
\def\eea{\end{eqnarray}}

\def\tt{{t}}

\def\half{{1\over 2}}
\def\ve{{\varepsilon}}

\def\tt{{t}}

\def\bmu{{\bar{\mu}}}
\begin{document}
\title{Metallic transport of hard core bosons}

\author{Sauri Bhattacharyya$^{1}$, Ayush De$^2$, Snir Gazit$^{2,3}$ and Assa Auerbach$^{1}$ }
\affiliation{$^1$Physics Department, Technion, 32000 Haifa, Israel\\
$^2$The Racah Institute of Physics, The Hebrew University of Jerusalem, Jerusalem 9190401, Israel\\
$^3$The Fritz Haber Research Center for Molecular Dynamics, The Hebrew University of Jerusalem, Jerusalem 91904, Israel}

  \date{\today }
\begin{abstract}   
Conductivities and Hall coefficients of two dimensional hard core bosons are calculated using the thermodynamic expansions of Kubo formulas.  At temperatures above the superfluid transition, the resistivity rises linearly and is weakly dependent on boson filling. The zeroth order Hall coefficient diverges toward zero and unit fillings, and reverses its sign at half filling. The correction terms, which are calculated up to fourth (Krylov) orders, do not alter this behavior. The high temperature thermal Hall coefficient is reversed relative to the electric Hall coefficient.
We discuss relevance of HCB transport to the metallic state of short coherence length superconductors.
\end{abstract}


\maketitle
\section{Introduction}
Two dimensional Hard Core Bosons (HCB) is a paradigmatic model of strongly interacting lattice bosons, and short coherence length superconductors.
HCB have modeled $^4$He  superfluid films~\cite{HCB-Matsubara}, cold optical-lattice bosons between Mott insulator phases~\cite{HCB-OL,Bloch-Simulations}, low capacitance Josephson junction arrays~\cite{Nandini,HCB-AA}, and  the superconducting fluctuations of cuprates~\cite{PBFM,HCB-Mihlin}. 

HCB density-temperature ($n\!-\!T$) phase diagram has been well explored by numerical simulations of the spin half quantum $XY$ model, using sign-free Quantum Monte Carlo (QMC) averaging~\cite{Loh-QMC,Ding-QMC1,Batrouni-QMC}. 
Below the Berezinskii~\cite{Ber}, Kosterlitz and Thouless~\cite{KT} (BKT)  temperature $T_{\rm BKT}(n)$, which is maximized at $n={1\over 2}$,   HCB  exhibit zero resistance and a finite superfluid stiffness, which can be explained by the {\em classical} $XY$ model. In a narrow regime of short range phase correlations above $T_{\rm BKT}(n)$, Halperin and Nelson (HN)~\cite{HN} showed that the resistivity rises due to proliferation of free vortices. 
In order to apply HN theory for the HCB resistivity, we need to know the ``normal metal'' resistivity, which is defined at a temperature where the  free vortices separation is reduced to the lattice constant scale.

However, the ``normal metal'' phase of HCB phase has received much less attention. This may be attributed to the inability of Boltzmann's  equation to properly account for hard core interactions and lattice Umklapp scattering, especially near half filling where the mean free path would be estimated as less than a lattice constant. 

The alternative is to directly compute real-frequency Kubo formulas, which faces severe numerical challenges.
Exact diagonalizations for the eigenstates (Lehmann) representation are
exponentially costly in  lattice size.  Analytic continuation of QMC data to real frequencies is ill-posed at frequencies lower than the temperature~\cite{Snirgaz}, which requires  the use of proxies for the DC limit~\cite{KivelsonAC}.

On the other hand, thermodynamic approaches to Kubo formulas~\cite{Mori,Zwanzig} enjoy the advantage of calculating equilibrium expectation values and static susceptibilities.
These are amenable to well established statistical-mechanics tools. For example, high temperature expansion, variational wavefunctions, and imaginary-time QMC avoid the high memory cost of exact diagonalizations, and the numerical pitfalls of analytic continuation.  

In this paper we apply two thermodynamic approaches to the HCB model.   (i) A continued fraction (CF) expansion~\cite{Viswanath} combined with a variational extrapolation of recurrents~\cite{RXX-PRB,Khait}. 
(ii) New thermodynamic summation formulas of Hall and thermal Hall coefficients~\cite{EMTPRL,EMT}. These formulas were previously  applied to  narrow gap semimetals~\cite{Abhisek}, and to the Hubbard~\cite{Dev-RXY}  and  tJ models~\cite{tJM} of strongly interacting electrons.   

\begin{figure}[ht!]
\begin{center}
\includegraphics[width=8cm]{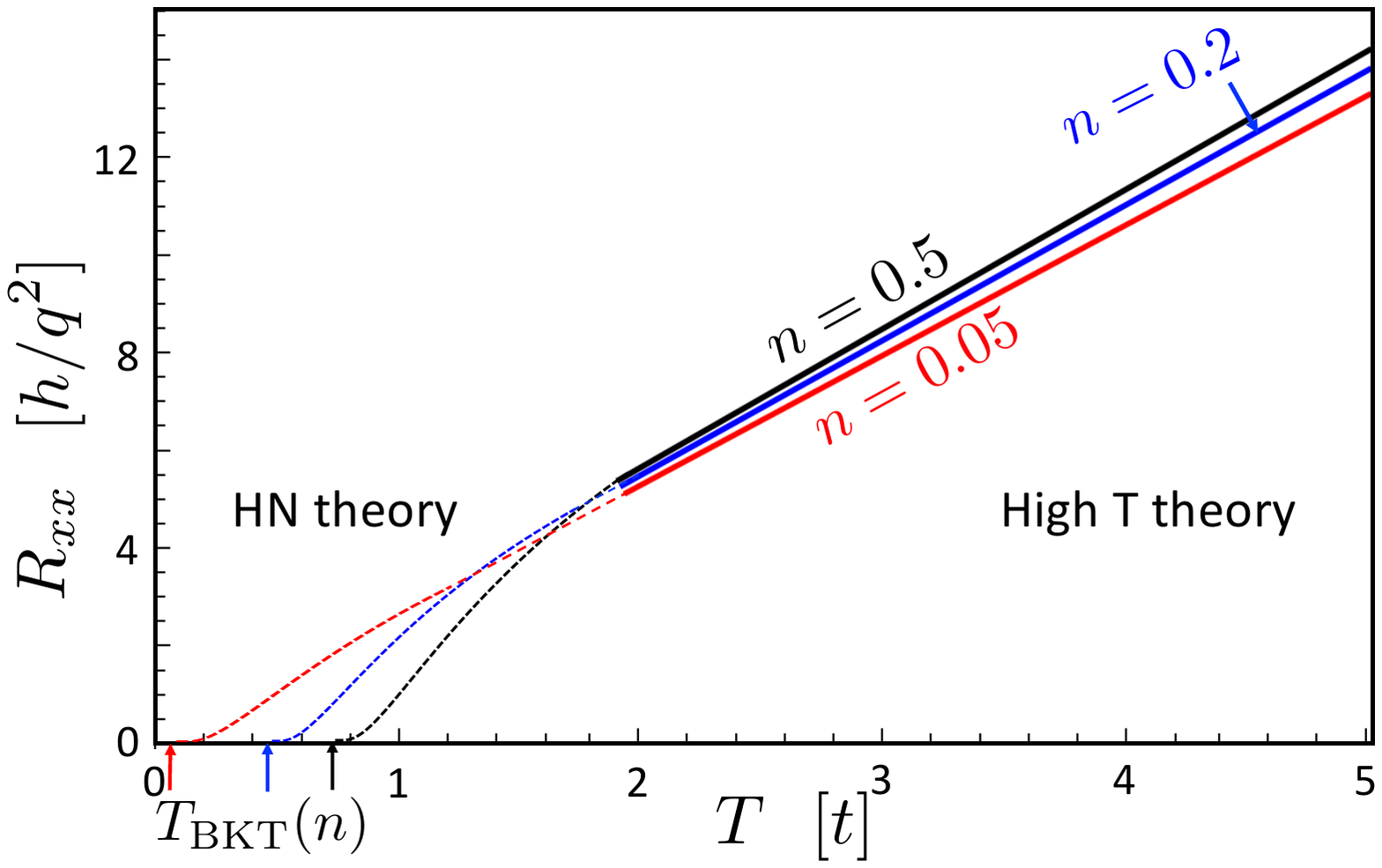}
\caption{Resistivity versus temperature of HCB  for several densities $0<n<{1\over 2}$. ($R_{xx}$ is symmetric for $n\to 1-n$).  
At high temperatures (solid lines)  $R_{xx}$ is calculated by Gaussian extrapolation of 5 lowest order recurrents, see Fig.~\ref{fig:DeltaCh-Ex}. The resistivities vanish exponentially (dashed lines) toward the superconducting transition at $T_{\rm BKT}\simeq 2.8 t n(1-n)$, following Halperin and Nelson's (HN) theory, Eq.~\ref{HN}. }
\label{fig:RXXvsT}
\end{center}
\end{figure}

Our main results are as follows. As shown in Fig.~\ref{fig:RXXvsT}, HCB resistivity rises with a linear slope whose value is insensitive to the density even down to  5\% filling.
This density independence is linked to a cancellation between the kinetic energy and the current relaxation  time.  

\begin{figure}[ht!]
\begin{center}
\includegraphics[scale=0.3]{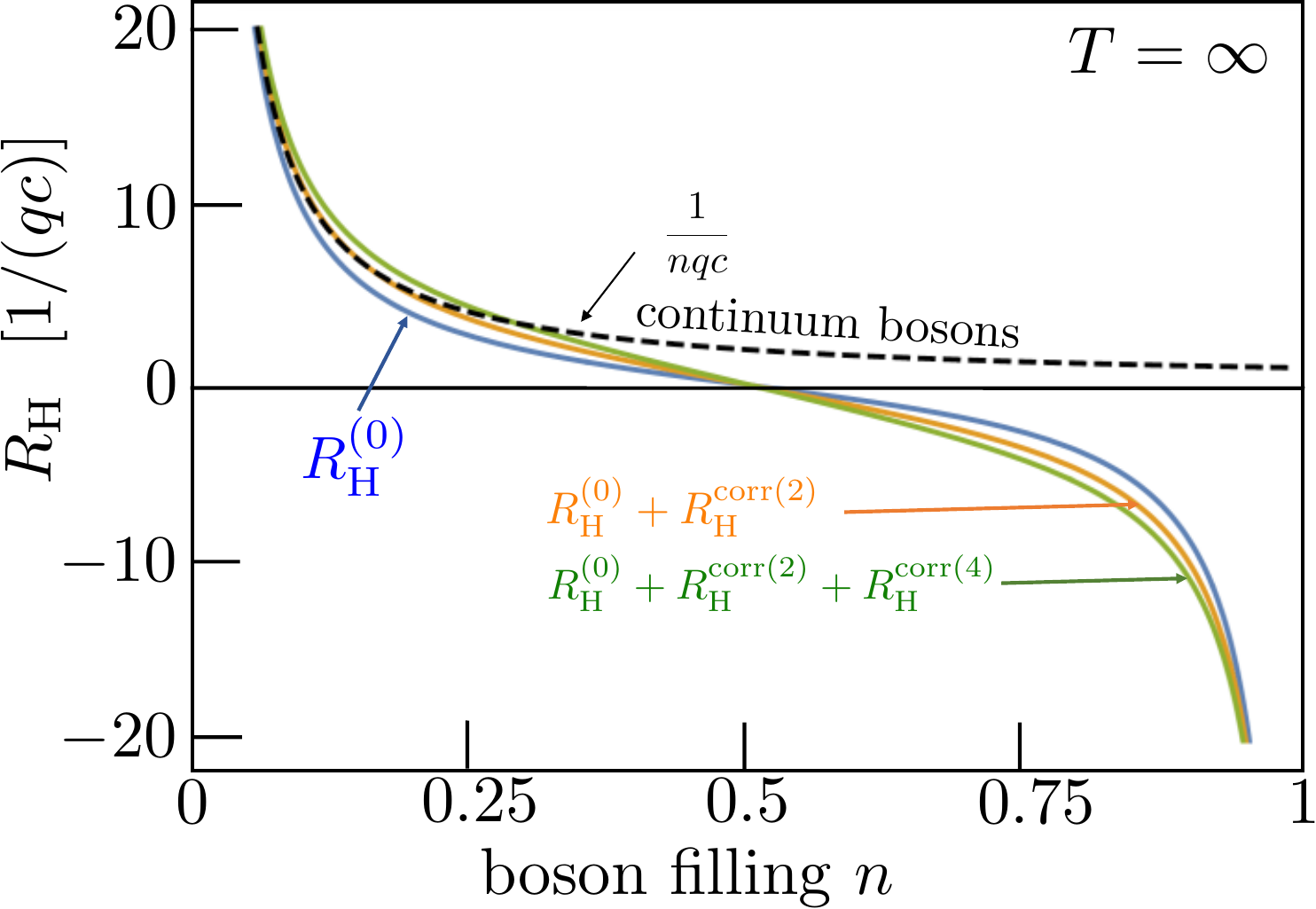}
\caption{Density dependent Hall coefficient $R_{\rm H}(n)$ of HCB on a square lattice at high temperature. $R_{\rm H}^{(0)}, R^{\rm corr}_{\rm H}$ are defined in Eqs.~(\ref{RH0}) and (\ref{Rcorr}).
Convergence  up to fourth order corrections is shown by the yellow and green curves. The  Hall sign change at half filling is a consequence of the the hard core interactions
on the lattice. 
}
\label{fig:RHwCorr}
\end{center}
\end{figure}

In Fig.~\ref{fig:RHwCorr} we plot the density dependent Hall coefficient, whose sign is  reversed at half filling relative to the Galilean result of continuum bosons.  The correction terms of the thermodynamic formula are calculated up to fourth order and shown to be relatively unimportant.
We also obtain the  thermal Hall coefficient, which is opposite in sign to the Hall effect, reflecting a ``cooling'' effect of the Hall current near half filling. 

This paper is organized as follows. In Section \ref{sec:HCB} the HCB model is defined. Section \ref{sec:RXX}  reviews the continued fraction expansion of the conductivity and the gaussian extrapolation of its recurrents.
Section \ref{sec:RH} describes 
the zeroth Hall coefficient term evaluated by a high temperature expansion and extended to low temperatures by   QMC.
Section \ref{sec:Rcorr} describes the calculation of the correction term up to fourth order. This calculation is crucially important for estimating the accuracy of the zeroth term. Section \ref{sec:RTH} calculates the  thermal Hall coefficient. We conclude by a summary and discussion of relevancy of our results to experiments, with particular emphasis on normal phase transport of short coherence-length superconductors such as cuprates. Experiments in cold atoms trapped in an optical lattice are also proposed.

\section{Hard Core Bosons Model} 
\label{sec:HCB}
The HCB creation and  density operators  at site $i$ are $\tilde{a}^\dagger_i, n_i$ respectively, which obey $[n_i,\tilde{a}_j^\dagger]= \delta_{ij}\tilde{a}_i^\dagger$.
The HCB constraint $(\tilde{a}^\dagger_i)^2=0$  is  faithfully  represented by spin half operators, $ \tilde{a}^\dagger_i \!\to\! S^+_i, n_i -\half \!\to\!  S_i^z$ (setting $\hbar\!\to\!1$).
On a square lattice with unit lattice constant, and total area $A$, we consider the gauged Hamiltonian of HCB, 
\be
H=- t \left(\sum_{\ij }e^{-i{q\over c} A_{ij}} S^{+}_{i}S^{-}_{j} + h.c\right) - \mu\sum_{i}S^{z}_{i} ,
\label{HCB} 
\ee 
where $\mu$ is the chemical potential and $q/c$ is the boson charge over velocity of light.  $\bA_{ij}=-{\br_{i}+\br_j\over 4} \times{\bf B}\cdot \br_{ij} $ introduces a uniform magnetic field $\bB \parallel \bz$.
The HCB charge polarizations, currents and magnetization operators are respectively represented by, 
\bea
&&\bP = q \sum_{i} ~ \br_i  S^z_i ,\quad \bj=i[H,{\bf P}]=\sum_{\ij} \bj_{ij}, \nonumber \\ 
&& j^\alpha_{ij}=- i q t (S^+_i S^-_j - S^-_i S^+_j)(r^\alpha_j-r^\alpha_i) ,\nonumber\\
&&M={1\over 4c} \sum_{\ij}  (\br_i +\br_j) \times \bj_{ij}.
\label{electric}
\eea
Here $\br_i$ denotes the position of site $i$.
The density dependent  $T_{\rm BKT}(n)$ for HCB on the square lattice~\cite{Ding-QMC1,Ding-QMC2,Harada} is
\be
T_{\rm BKT}(n) \simeq 2.8 t n(1-n).
\ee

\section{DC Resistivity} 
\label{sec:RXX}
The DC longitudinal conductivity is given by the CF expansion,
\be
\sigma^{\rm dc}_{xx} = \chi_{\rm csr}\lim_{\ve\to 0^+} \Im {1\over -i\varepsilon+{\Delta_1^2\over  - i\ve +{\Delta^2_2\over
\ddots }}} 
\label{CF}
\ee
where $\chi_{\rm csr}$ is the conductivity sum rule (CSR),
\be
\chi_{\rm csr}= {1\over A} \Im \langle [P^x,j^x]\rangle 
\ee
and $\Delta_k,k=1,\ldots,k_{\rm max} $ are the calculated recurrents. 
The recurrents are obtained from
the conductivity moments, which are calculated as thermodynamic expectation values,
\be
\mu_{2k} =   - {1\over A}  \Re \langle [j^x,\overbrace{[H,[H,\ldots [H}^{2k-1},j^x]]]\ldots]\rangle,~
\ee
$j^x$ is the uniform current  in the $x$ direction defined in Eq.~(\ref{electric}).

The CSR and the $\mu_{2k}$ are expanded in powers of inverse temperature $\beta$ as described in Appendix \ref{App:HiT}. The CSR up to order $\beta^3$ is given by,
\bea
\chi_{\rm csr} &=&   \beta q^2 t^2n \left(1-n\right) \Big(2 \nonumber\\
&&~+\frac{(\beta t)^2}{3}\left(1-n\right)^2\left(-3 + 10 n (1-n))\right)\Big).
\label{CSR-c}
\eea

\begin{figure}[h]
\begin{center}
\includegraphics[width=7cm,angle=0]{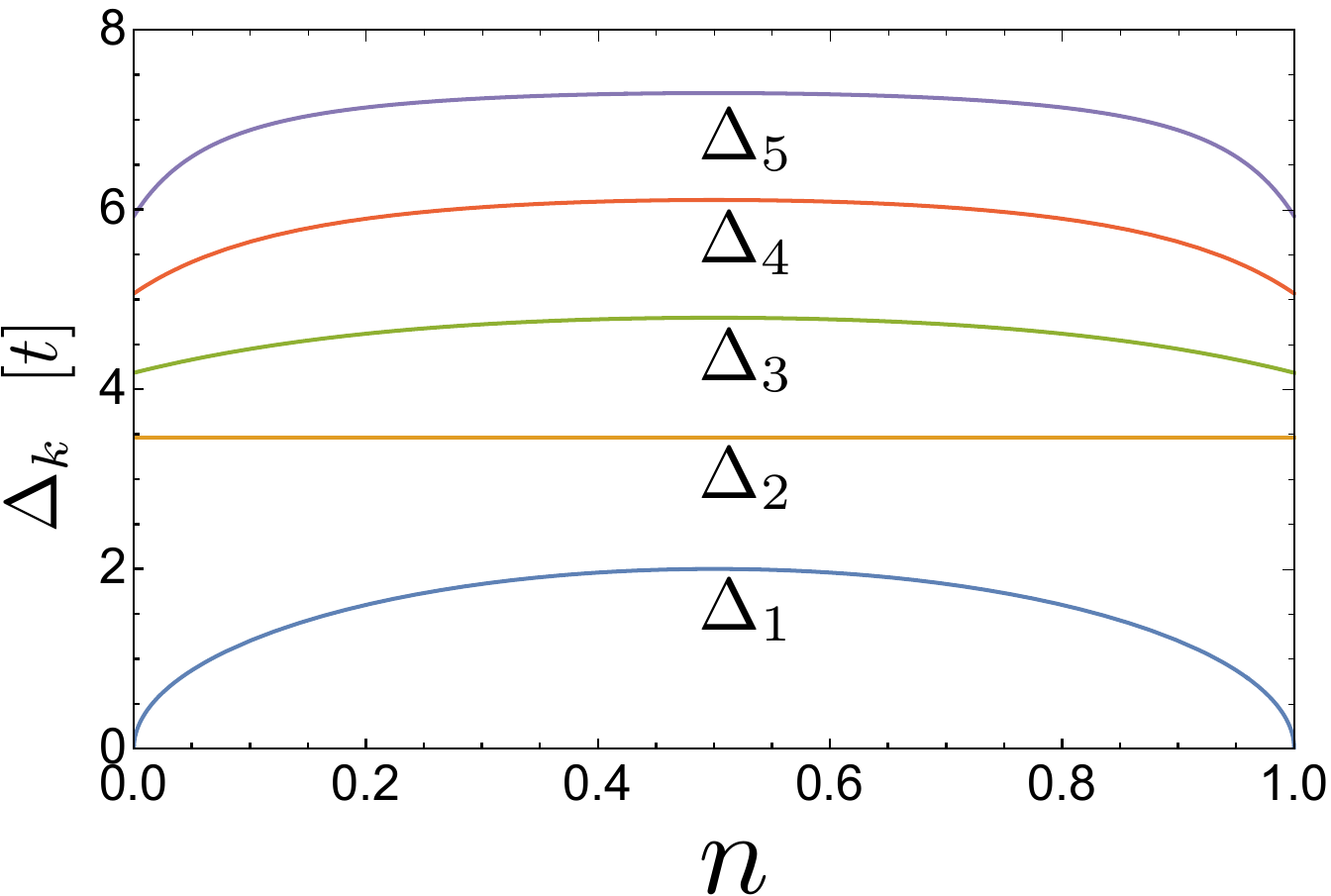}
\caption{Density dependence of the five lowest recurrents. The behavior of $\Delta_1$ in Eq.~(\ref{Delta1}) is largely responsible for the weak density dependence of the resistivity slopes in Fig.~\ref{fig:RXXvsT}. }
\label{fig:Deltas}
\end{center}
\end{figure}

The five lowest normalized moments $\bmu_{2k}=\mu_{2k}/\chi_{\rm csr}$ are shown in  Table ~\ref{table:bmu}. They involve traces over many operators which are evaluated  by symbolic multiplication, as explained in Appendix \ref{App:Auto}. 
\begin{table}[t]
\begin{tabular}{c p{8cm}}
\hline
$2k$ & $\bmu_{2k}/(\beta  t^{2k})$ \\
\hline
2 & $ 16n(1-n)$ \\
4 & $ 64 n(1-n) n (3 + 4 n - 4 n^2) $ \\
6 & $  32 n(1-n)  (177 + 356 n - 356 n^2)$ \\
8 & $  128 n(1-n)   (1979 + 7520 n - 10432 n^2 + 7040 n^3 - 6560 n^4 + 
   3648 n^5 - 1216 n^6) $ \\
10 & $  128 n(1-n)  (119200 + 856443 n - 1386927 n^2 + 1358488 n^3 - 
   1459972 n^4 + 1040272 n^5 - 519088 n^6 + 147712 n^7 - 36928 n^8)$ \\
\hline
\end{tabular}
\caption{Normalized conductivity moments $\bmu_{2k}=\mu_{2k}/\chi_{\rm csr}$ as functions of density $n$ at leading order in $\beta$.}
\label{table:bmu}
\end{table}
The  recurrents $\Delta_k$ in Eq.~(\ref{CF}) are obtained from the 
normalized moments by the algebraic relations given in Appendix \ref{App:recurrents}. The five lowest order recurrents are,
\bea
\Delta_1^2 &=& \bmu_2,\nonumber\\
\Delta^2_2 &=& {\bmu_4\over \bmu_2} - \bmu_2,\nonumber\\
 \Delta_3^2 &=& {\bmu_4^2  -\bmu_2\bmu_6\over\bmu_2^3 - \bmu_2\bmu_4 },\nonumber\\
 \Delta_4^2 &=& \bmu_2 {(\bmu_4^3 + \bmu_6^2 + \bmu_2^2\bmu_8 - \bmu_4 (2\bmu_2 \bmu_6 + \bmu_8))\over
(\bmu_2^2 - \bmu_4 ) (\bmu_2 \bmu_6 - \bmu_4^2 )},\nonumber\\
\Delta_5^2 &=& {  \bmu_4-\bmu_2^2  \over  \bmu_4^2 - \bmu_2 \bmu_6  }\nonumber\\
 &&\times {    \bmu_{10} (\bmu_2 \bmu_6 -\bmu_4^2 ) -\bmu_6^3 + 2 \bmu_4 \bmu_6 \bmu_8 - 
   \bmu_2 \bmu_8^2   \over   \bmu_4^3 + \bmu_6^2 + \bmu_2^2 \bmu_8 - 
   \bmu_4 (2 \bmu_2 \bmu_6 + \bmu_8)  }.
\label{recurrents-moments}
\eea
In Fig.~\ref{fig:Deltas}, the  recurrents of Eqs. \ref{recurrents-moments}  are plotted  for densities $0<n<1$.

For the square lattice, the leading high temperature terms are $\chi_{\rm csr}\propto \beta$, and $\Delta_k=\cO(\beta^0)$. Thus conductivity goes as $\beta$, and the  resistivity is asymptotically linear. 
\begin{figure}[ht!]
\begin{center}
\includegraphics[width=10cm]{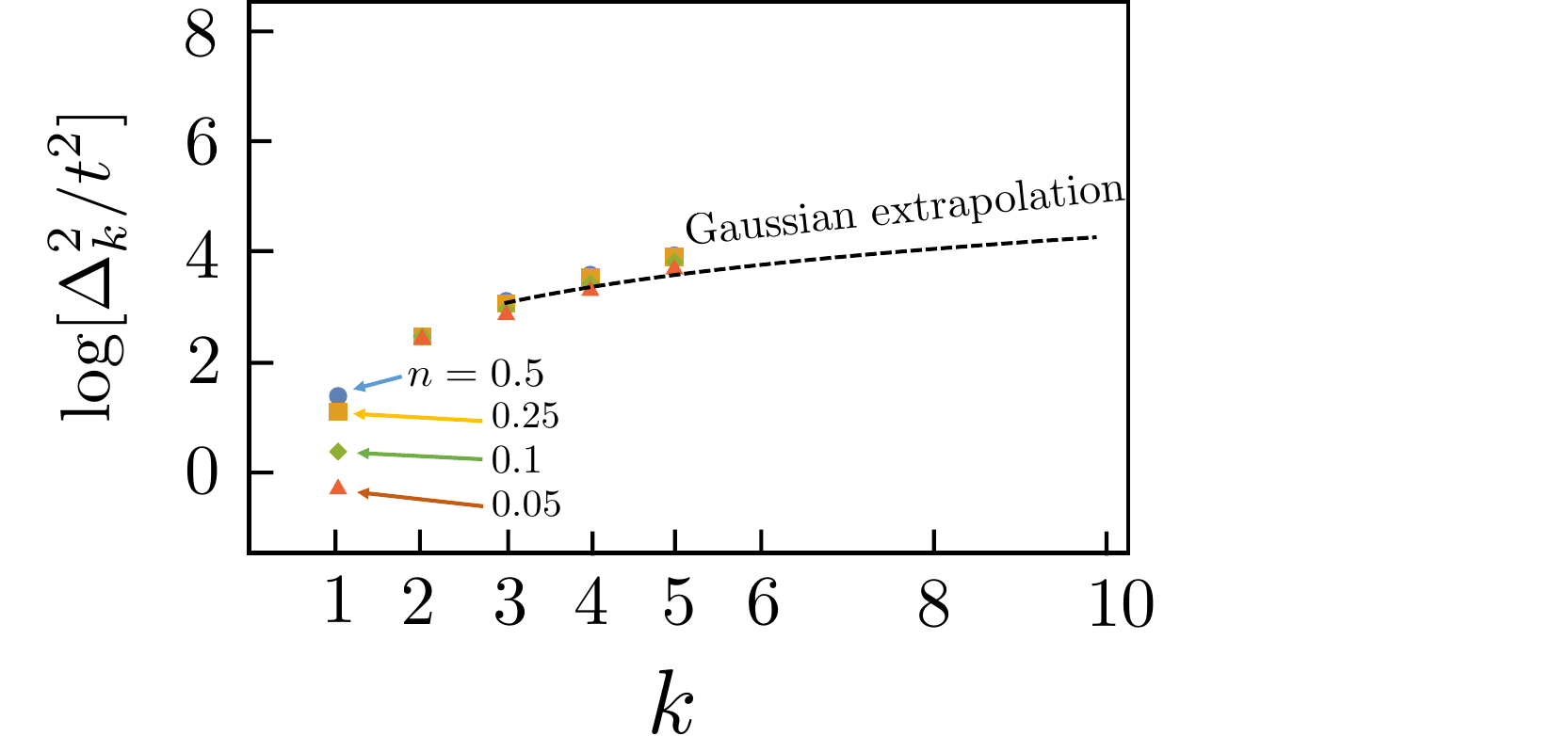}
\caption{High temperature conductivity recurrents of HCB.  Solid circles are calculated recurrents, dashed line  describes the Gaussian extrapolation based on the third to the fifth calculated recurrents at density $n=0.3$. }
\label{fig:DeltaCh-Ex}
\end{center}
\end{figure}

In Fig.~(\ref{fig:DeltaCh-Ex}) the five lowest recurrents are depicted on a log-linear plot, for densities varying from 5\%-50\%. We note that the first  recurrent is,
\be
\Delta_1^2 = 16 n(1-n)t^2.
\label{Delta1}
\ee
We also notice  in Figs.~\ref{fig:Deltas} and  \ref{fig:DeltaCh-Ex} that $\Delta_{k>1}$  exhibit a weaker density dependence.

 \subsection{Gaussian termination function}
 \label{sec:Gauss}
Having calculated  the recurrents
$\Delta_1,\Delta_2\ldots \Delta_{k_{\rm max}}$, the CF in Eq.~\ref{CF} depends on the imaginary part of an unknown termination function $G''_{k_{\rm max}+1}=\lim_{\ve\to 0} \Im G_{k_{\rm max}+1}(i\ve)$,
\be
\sigma_{xx}^{\rm dc}= \chi_{\rm csr} {1\over  {\Delta_1^2\over { \vdots\over    \Delta^2_{k_{\rm max}}  G''_{k_{\rm max}+1}(0^+)}}} 
\label{TF}
\ee
The variational extrapolation of recurrents (VER) method determines
$G_{k_{\rm max}+1}$ as follows.

 For the HCB model at $n=0.5$, VER yielded good agreement~\cite{RXX-PRB} between the extrapolated high temperature conductivity
 and the Kubo formula  computed by exact diagonalization.
The choice of a gaussian variational function,
\be
F(\omega) = {\sqrt{\pi}\over \Omega} \exp\left( -{\omega^2\over \Omega^2}\right),~~~
\ee
proved adequate. Its variational recurrents are,
\be
\bar{\Delta}_k^2 =k {\Omega^2 \over 2}. 
\ee
In Fig.~\ref{fig:DeltaCh-Ex} we show a good fit of the dashed line  to the computed recurrents $\Delta_3,\Delta_4, \Delta_5$, for the variartional choice
\be
\Omega^2  =  {2\over  3} \Delta_3^2 .
\ee
The corresponding termination function $\bar{G}''_6$ is determined by,
\be
\Im \bar{F}(\omega\to 0^+) =   { \bar{\Delta}^2_2 \bar{\Delta}^2_4  \over \bar{\Delta}^2_1 \bar{\Delta}^2_3   \bar{\Delta}_5^2 G''_6} = \sqrt{3\pi\over 2}  {1\over \Delta_3} .
\ee
Hence,
\be
\bar{G}''_6=   {8\over 5} \sqrt{2\over 3\pi}  {1\over \Delta_3} .
\ee
Thus we obtain,
\be
\sigma^{\rm dc}_{xx}  \simeq {\chi_{\rm csr}(n) \over \Delta_1^2(n) }\times  \sqrt{3\pi\over 2} \times {5 \Delta_2^2 \Delta_4^2  \over  8 \Delta_5^2 \Delta_3} .
\label{SxxDC}
\ee

Physically,  $\chi_{\rm csr}$ measures the HCB kinetic energy and $\Delta_1^2/\Delta_2$ describes the current dissipation rate. The  factors of $n(1-n)$ factors cancel out between $\chi_{\rm csr}$ and $\Delta_1^2$.
This can explain the approximate  density independence of the conductivity found by Gaussian extrapolation in Section \ref{sec:Gauss} and shown in Fig.~\ref{fig:RXXvsT}.

We ignore the weak density dependence of the leading order in $\beta$ we invert
$\sigma_{xx}^{\rm dc}$ to obtain the high temperature resistivity, 
\be
R_{xx} \simeq 3.3  ~ {h \over q^2   } {T\over t}.
\label{RT}
\ee 

Here we have not calculated the next order $\beta^3$ correction of $\sigma_{xx}^{\rm dc}$ as a function of density. We note that Ref.~\cite{RXX-PRB} found for $n=0.5$ a relative correction
$ - 0.75 \left({t\over T}\right)^2 $, which at $T=4 T_{\rm BKT}$  is less than $10\%$.

\section{Hall coefficient} 
\label{sec:RH}
 The Hall coefficient formula~\cite{EMT} contains two contributions, 
\be
R_{\rm H}=R_{\rm H}^{(0)}+R^{\rm corr}_{\rm H}.
\label{RH}
\ee
The first term is given by
\be
R_{\rm H}^{(0)} = {\chi_{\rm cmc}\over \chi_{\rm csr}^2 }.
\label{RH0}
\ee
The current-magnetization-current (CMC) susceptibility expanded to order $(\beta t)^2$
is
\bea
\chi_{\rm cmc} &=& 2 (j^y,[M,j^x])\nonumber\\
&=& 4\frac{\beta^2  q^3 t^4}{c}(1-2n)n(1-n) \Big(1  \nonumber\\
&&~-(\beta t)^2  (1 - 3 n(1-n))\Big).
\label{CMC-c}
\eea

\begin{figure}[ht!]
\begin{center}
\includegraphics[width=8cm]{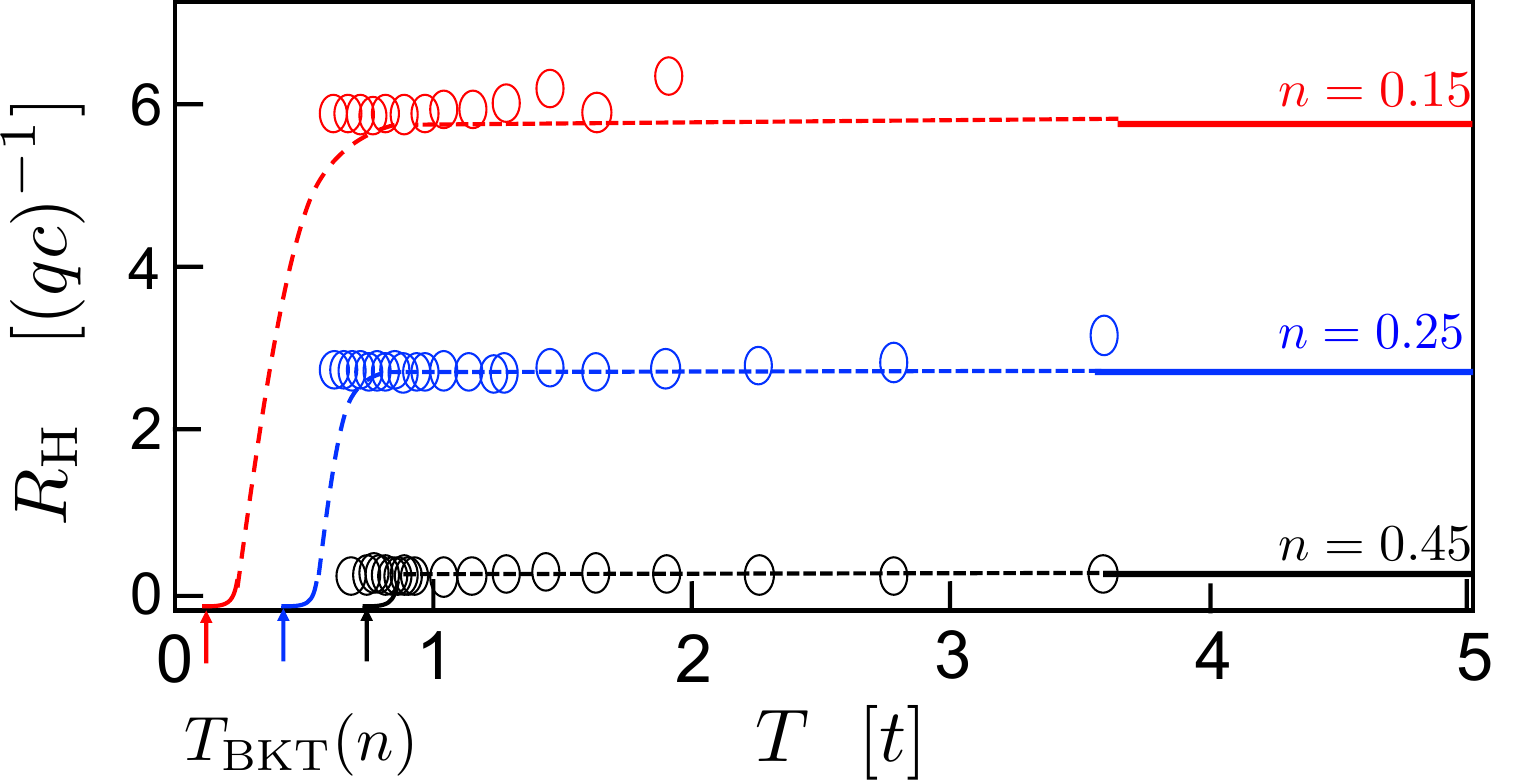}
\caption{Metallic Hall resistivity of HCB.  
High temperature asymptotes $R_{\rm H}^{(0)}$ of Eq.~(\ref{RH0-highT}) (solid lines), and  lower temperature extension by QMC calculations (circles). Dashed lines interpolate between $R_{\rm H}^{(0)}(T)  $  and HN theory near $T_{\rm BKT}(n)$, see Section \ref{sec:HN} }
\label{fig:RHvsT}
\end{center}
\end{figure}

Note that $\chi_{\rm csr}$ ($\chi_{\rm cmc}$) is ``particle-hole'' symmetric (antisymmetric) under $n\to 1-n$. Thus we obtain,
\be
R_{\rm H}^{(0)}=  \frac{1}{qc}\left(\frac{2n-1}{n(n-1)} + \frac{2}{3}(\beta t)^2\left(n-\frac{1}{2}\right)\right). 
\label{RH0-highT}
\ee
We note that at low density, the Hall coefficient recovers the continuum Galilean invariant result
$R_{\rm H}^{(0)} \to (nqc)^{-1}$. Near half-filling, $R_{\rm H} \!\sim \!  -8({n-{1\over 2})/(qc)}$ reflecting the effects of lattice Umklapp and hard core scattering.

Eq.~(\ref{RH0-highT}) was extended to lower temperatures numerically by a path-integral based QMC 
for bosonic lattice models~\cite{Motoyama} which are devoid of a sign problem. We studied  $24 \times 24$ size lattices, which was sufficiently large for expectation values at temperatures higher than $T_{\rm BKT}$. The number of Monte Carlo sweeps was 
$\sim 5\times10^{6}$. $\beta$ was subdivided into intervals $\epsilon=\beta/N_\epsilon, N_\epsilon=10$.
Off-diagonal  operators (e.g. $S^{+}_{i}S^{-}_{j}$, $S^{+}_{i}S^z_jS^{-}_{k}$) were calculated using using worm-type updates. 
In Fig.~\ref{fig:RHvsT}, the solid lines are the analytical results of Eq.~(\ref{RH0-highT}), while the QMC data are depicted by  open circles. We see that the Hall coefficients above the HN regime, rapidly saturate  to their high temperature limit. 

\subsection{The Correction term} 
\label{sec:Rcorr}
The correction term in Eq.~(\ref{RH}) contains a sum of susceptibilities of Krylov operators: 
\bea
&&R_{\rm H}^{\rm corr}= {1\over \chi_{\rm csr}}  \sum_{i,j=0}^\infty R_i R_j  (1 -\delta_{i,0} \delta_{j,0})M''_{2i,2j}\nonumber\\
&&M_{2i,2j}=  \Im \left( \langle 2i;y |  \cM |2j; x \rangle - \langle 2i; x | \cM |2 j;y\rangle \right)  ,\nonumber\\
&&R_{i>1} = \prod_{r=1}^i \left(-{\Delta_{2r-1}\over \Delta_{2r} } \right),\quad R_0 =1.
\label{Rcorr}
\eea
The conductivity recurrents $\Delta_n$  are evaluated in Fig.~\ref{fig:Deltas}.

The unnormalized matrix elements $M''_{nm}$  of Eq.~(\ref{Rcorr}) involve   
susceptibilities of the form
\be
\tilde{M}''_{nm}=(\cL^n j^y, [M,\cL^m j^x]).
\ee 
Unnormalized hyperstates  can be expanded in terms of the orthonormal Krylov bases by the Gram-Schmidt matrix $K$,
\be
\cL^{n} j^\alpha = \sum_{k'=0}^{k} K_{n,k'} |k',\alpha\rangle,
\ee
where $\langle k,\alpha|k',\alpha'\rangle=\delta_{kk'}\delta_{\alpha\alpha'}$.
The factors $R_i$ involve a finite number of  recurrents $\Delta_1,\ldots \Delta_{2i}$, which also determine $K^{-1}$ up to the same order.

The simplicity of the Hamiltonian permits a calculation of up to fourth order corrections,
\be
R_i R_j M''_{2i,2j},\quad i,j=0,2,4.
\ee
The magnetization matrix elements involved  traces over up to $\sim 10^7$ operator products. The hypermagnetization matrix elements $M''_{nm}$, which are listed in Table~\ref{table:Mcab0} of Appendix~\ref{App:Corr}.

Since the Hall coefficient is finite for a metal, the
summation over all the higher order corrections must converge. The  corrections  to $R_{\rm H}^{(0)}$ up to fourth order are depicted in Fig.~\ref{fig:RHwCorr}. 
We see that these corrections do not qualitatively change the zeroth term's behavior especially near the densities $n=0,1,{1\over 2}$, although they  converge slower around intermediate densities $n= 0.25,0.75$. 

In conclusion, $R^{(0)}_{\rm H}$ appears to be a qualitatively correct approximation at high temperatures. $T>t$.

In the lower temperature regime  $R_{\rm H}^{(0)}$, as evaluated by QMC, appears to be blind to the onset of long range phase correlations and vortex fluctuations as described by HN theory. This implies imply that
that the correction term should grow  in magnitude as $T\to T_{\rm BKT}$,
and approach  
$R_{\rm H}^{\rm corr} \to -R_{\rm H}^{(0)}$,  to comply with the onset of superconductivity.

\section{Matching  Halperin-Nelson Theory} 
\label{sec:HN}
Halperin and Nelson (HN) described a narrow fluctuation region just above the two dimensional  superconducting transition at $T_{\rm BKT}$.
In that regime, the resistivity tensor is dominated by the exponentially small density of free vortices   which vanishes  $T_{\rm BKT}$.
At higher temperatures, the vortex density increases such that they cease to be well defined degrees of freedom. According to  (HN) theory~\cite{HN},  
\bea
R^{\rm HN}_{\alpha\beta} &\simeq& 2.7 R^{\rm n}_{\alpha\beta}(T)\left({\xi_+\over \xi_c}\right)^{-2} \nonumber\\
&=&  2.7 R^{\rm n}_{\alpha\beta}(T)~ \exp\left(-2b\left( {T_{\rm BKT}\over T-T_{\rm BKT} } \right)^{1\over 2} \right).
\label{HN}
\eea
$\xi_+$ is the BKT correlation length, and $\xi_c$ is of the order of the HCB lattice constant and $b\simeq 1$.  

For the HCB model, the ``normal state'' resistivities $R^{\rm n}_{xx}(T)$ and  $R^{\rm n}_{yx}(T)$,
is taken from our Eq.~(\ref{RT})  (\ref{RH0-highT}), respectively.

We use these values to plot the crossovers from HN theory Eq.~(\ref{HN}) to higher temperatures as dashed lines in Figs.~\ref{fig:RXXvsT} and \ref{fig:RHvsT}.

\section{Thermal Hall coefficient} 
\label{sec:RTH}
The heat currents are defined using the energy polarization $\bP_{\rm E}$, and charge current $\bj$,
\bea
&&\bj_{\rm Q}= i[H,\bP_{\rm E}] - {\mu\over q} \bj\nonumber\\
&&\bP_{\rm E}=  \sum_{\ij} ~ { \bx_i + \bx_j \over 2}  ~ h_{ij} ,\nonumber\\
&&h_{ij} = -t \left( S^{+}_{i}S^{-}_{j} + S^{-}_{i}S^{+}_{j} \right).  
\label{jQ}
\eea

\begin{figure}[ht!]
\begin{center}
\includegraphics[width=8.5cm]{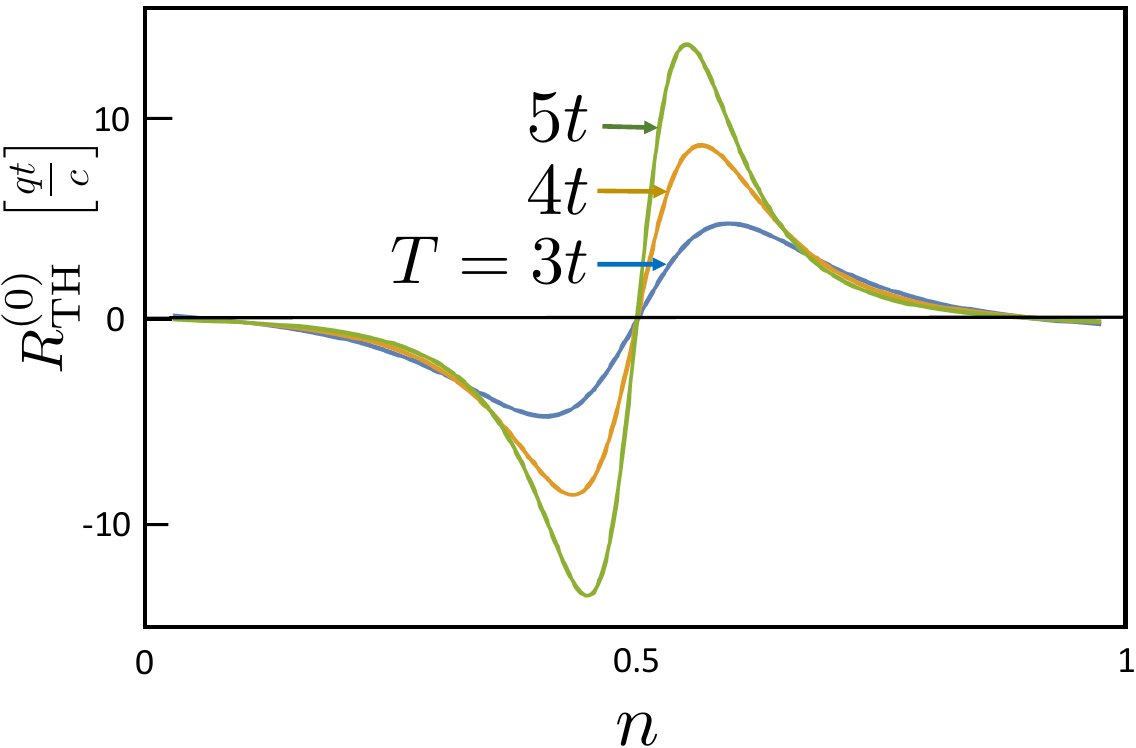}
\caption{Thermal Hall coefficient of HCB versus density at three high temperatures.  }
\label{fig:RTH}
\end{center}
\end{figure}

The zeroth thermal Hall coefficient is given by~\cite{EMT},
\be
R_{\rm TH}^{(0)}={ \chi_{\rm cmc}^Q\over \beta (\chi_{\rm csr}^Q)^2},
\label{RTH}
\ee
where the heat current susceptibilities~\cite{ShastrySR} are decomposed into  energy-energy (EE) currents, energy-charge (EC) and the previously introduced charge-charge  susceptibilities $\chi_{\rm cmc},\chi_{\rm csr}$:
\bea
&&\chi_{\rm csr}^{\rm Q}\!=\! (j_{\rm Q}^x|j_{\rm Q}^x) \!=\!  \chi_{\rm csr}^{\rm ee} - {2\mu\over q}\chi_{\rm csr}^{\rm ec}+\left({\mu\over q} \right)^2 \chi_{\rm csr},\nonumber\\
&&\chi_{\rm cmc}^{\rm Q} \!=\!  2(j_{\rm Q}^y|M|j_{\rm Q}^x) \!=\!  \chi_{\rm cmc}^{\rm ee}\!-\!  {2\mu\over q}\chi_{\rm cmc}^{\rm ec}\!+\! \left({\mu\over q} \right)^2\chi_{\rm cmc}.\nonumber\\
\label{chi-Q}
\eea
The energy-energy susceptibilities  are
\bea
\chi^{\rm ee}_{\rm csr} &=& \beta t^4  n (6  - 14 n + 16 n^2 - 8 n^3) ,
\nonumber \\
\chi^{\rm ee}_{\rm cmc} &=& {q \beta^2 t^6 \over c} \left( 24 n - 112 n^2 + 208 n^3 - 200 n^4 + 80 n^5 \right).\nonumber\\
\label{EE}
\eea
The energy-charge  susceptibilities are 
\bea
\chi_{\rm csr}^{\rm ec}&=&{3q\beta^2 t^4\over 2} (2n-1) (1 - (2n-1)^2)  , \nonumber\\
\chi_{\rm cmc}^{\rm ec} &=& - {q^2\beta t^4\over 2}  (1 - (2n-1)^4).\nonumber\\
\label{EC}
\eea
At high temperatures, the chemical potential $\mu$ is given by
\be 
e^{\beta\mu}={n\over 1-n}\Rightarrow \mu(\beta) =\beta^{-1} \log\left({n\over 1-n}\right).
\label{muT}
\ee

In Fig.~\ref{fig:RTH} we plot the zeroth thermal Hall coefficient defined in  Eq.~(\ref{RTH}) for three high temperatures. $R^{\rm corr}_{\rm TH}$, and higher order in $\beta$ corrections are not included here.

Since $\mu\sim 4 \beta^{-1}\left( n-{1\over 2}\right)$, by Eq.~(\ref{EC}), the leading contribution at high temperature near half filling is coming from the contribution of
\be
-2{\mu \over q}\chi^{\rm ec}_{\rm cmc}\propto -\left(n-{1\over 2}\right).
\ee 
The other $\cO(\beta^0)$ term  goes as $ \mu^2 \chi_{\rm cmc}\sim \left( n-{1\over 2}\right)^3$,
which is subdominant near half filling.
Therefore $R_{\rm TH}$ near half filling ends up having the opposite sign to the charge Hall coefficient. This (a-priori unexpected) result can be viewed as a  ``cooling'' effect of the charge Hall current on the transverse temperature difference.

\section{ Discussion and Summary}
\label{sec:Summary}
The  resistivity, Hall and thermal Hall coefficients were calculated for the metallic phase  of HCB at temperatures above the HN superconducting fluctuations regime.

Near half filling, HCB cannot be reasonably approximated by weakly interacting quasiparticles. The model includes quantum mechanical effects of lattice periodicity combined with strong local constraints of no-double occupancies. The resulting  conductivities can be classified as  ``non-Fermi liquid'' metallic transport. The strong interactions strongly affect the moments of conductivity and the resulting magnitude of the linearly increasing resistivity.

The sign reversal of the Hall coefficient at half filling is also understood as an effect of strong repulsive interactions. HCB differ from continuum bosons, which are not expected to reverse their Hall sign as a function of filling~\cite{Huber}.

Incidentally, we note that HCB and the tJ model of electrons~\cite{tJ} are somewhat similar in their proximity to Mott insulators.  
The Hall sign the tJ model also diverges toward the Mott phase, and exhibits a sign reversal relative to
weakly interacting quasiparticles.

{\em Experiments} --  HCB  may be realized in low capacitance gated  Josephson arrays at incommensurate fillings~\cite{Charlie}.
Measurements of temperature and density dependent resistivity and Hall coefficient could be compared to Figs.~\ref{fig:RXXvsT},\ref{fig:RHvsT}, and \ref{fig:RTH} respectively.

The phase diagram of layered superconducting cuprates has been described by the classical (highly anisotropic) layered $XY$ model~\cite{Lemberger,HCB-Mihlin}. This description is supported by Uemura's empirical relations between between superfluid stiffness 
and $T_{\rm c}$~\cite{Uemura,Bozovic}. It is therefore natural to  study  Eq.~(\ref{HCB}) in order  to understand dynamical responses slightly above $T_{\rm c}$. 

Systematic studies~\cite{Asban,Comm:Homes} have found an empirical proportionality between the resistivity slopes of {\em optimally doped} cuprates, and the inverse zero temperature superconducting stiffness. It is
quite  natural for  HCB which are goverened by the single energy scale $t$ governing both quantities.
The  linearly rising resistivity in Fig.~\ref{fig:RXXvsT} suggests that one should consider that a significant part of the transport current, slightly above $T_{\rm c}$, may be effectively carried by HCB that describe preformed (tightly bound) Cooper pairs. This may resolve some of the 
``bad metal'' conundrum~\cite{badmetals} in which the large magnitude of resistivity seems inconsistent with well defined Fermi liquid quasiparticles.

Finally, cold bosonic atoms trapped on optical lattices can serve as  platforms for measuring HCB conductivities using time dependent potentials. A Hall effect  can be induced by  artificial gauge fields~\cite{Spielman-SFHall,Spielman-Gauge,Bloch-SingleSite}.
We hope our results will motivate such experiments.

{\em  Acknowledgements} -- We thank Anna Keselman and Abhisek Samanta for consultations in the 
numerical implementation. The QMC calculations were done with the help of the DSQSS package and S.B. sincerely thanks Naoki Kawashima and 
Yuichi Motoyama for their kind help. AA acknowledges the Israel Science Foundation
(ISF) Grant No. 2081/20. S.G. acknowledges support from the Israel Science Foundation (ISF) Grant no. 586/22 and the US–Israel Binational Science Foundation (BSF) Grant no. 2020264. This work was
performed in part at the Aspen Center for Physics,
which is supported by National Science Foundation grant
PHY-2210452, and at the Kavli Institute for Theoretical
Physics, supported by Grant Nos. NSF PHY-1748958,
NSF PHY-1748958 and NSF PHY-2309135. 

\appendix

\section{Calculating recurrents from moments}
\label{App:recurrents}
The moments are related to the recurrents by the matrix equation~\cite{RXX-PRB}, 
\be
\mu_{2k}= \chi_{\rm csr} (L^{2k})_{00},
\label{bmu-L}
\ee
where the tridiagonal Liouvillian matrix is defined as, 
\be 
L_{ij}[\Delta] = \left( \begin{array} {cccccc}
0&\Delta_1&0&0&0&\ldots \\
\Delta_1&0&\Delta_2&0&0&\ldots \\
0&\Delta_2&0&\Delta_3&0&\ldots \\
0&0&\Delta_3&0&\Delta_4&\ldots \\
0&0&0&\Delta_4&0\ldots \\
0&\vdots &\vdots&\vdots&\vdots&\ldots \\
\end{array}\right)
\ee
Taking  even powers of $L$  and evaluating their $(00)$ matrix elements, yields relations between $\mu_{2k}$ and   the preceeding recurrents $\Delta_1,\ldots \Delta_k$, which are symbollically solved to obtain Eqs.~(\ref{recurrents-moments}).

\section{High temperature expansions of  susceptibilities}
\label{App:HiT}
We use the XY model representation, Eq.~(\ref{HCB}) to derive the following high temperature results.
In the grand canonical ensemble, the mean magnetization for the XY model $m=2n-1$, where $0\le n\le 1$ is the HCB density per site, can be imposed at infinite temperature by a product density matrix $\rho_0(\delta)$, with a fugacity parameter $\delta$:
\be 
\rho_0(\delta)= \prod_i  \left(  {1+\delta \over 2} |\uparrow \rangle\langle \uparrow | + {1-\delta \over 2} |\downarrow\rangle\langle \downarrow|  \right),
\label{Tr1}
\ee 
where the average magnetization at infinite temperature is
\be
m(0)=\langle 2S^z_i\rangle_{\beta=0} \equiv  \delta .
\label{Tr2}
\ee
Expectation values are,
\bea
\Tr \rho_0 S^x_i&=&0,\nonumber\\
\Tr \rho_0 S^y_i&=&0,\nonumber\\
\Tr \rho_0 S^z_i&=&\frac{1}{2}\delta .
\label{Tr3}
\eea

At finite temperature, we expand the average magnetization $m(\delta)$ to second order in $\beta$,
\bea
m &=& {\Tr \rho_0(\delta) e^{-\beta H}2S^z_i  \over \Tr   \rho_0(\delta_0) e^{-\beta H} } = \delta + 
{\beta^2\over 2} \Tr \rho_0(\delta) H^2 (2S^z_i - \delta )\nonumber\\
&=& \delta - \beta^2 \delta (1- \delta^2) + \cO(\beta^4).
\label{mag-corr}
\eea
Eq.~(\ref{mag-corr}) allows us to evaluate the magnetization dependence of a $\delta$-dependent susceptibility by
\be
\chi(\delta, \beta) = \chi (m+ \beta^2 \delta (1- \delta^2) ,\beta).
\label{corr}
\ee
 
\begin{figure}[h]
\begin{center}
\includegraphics[width=7cm,angle=0]{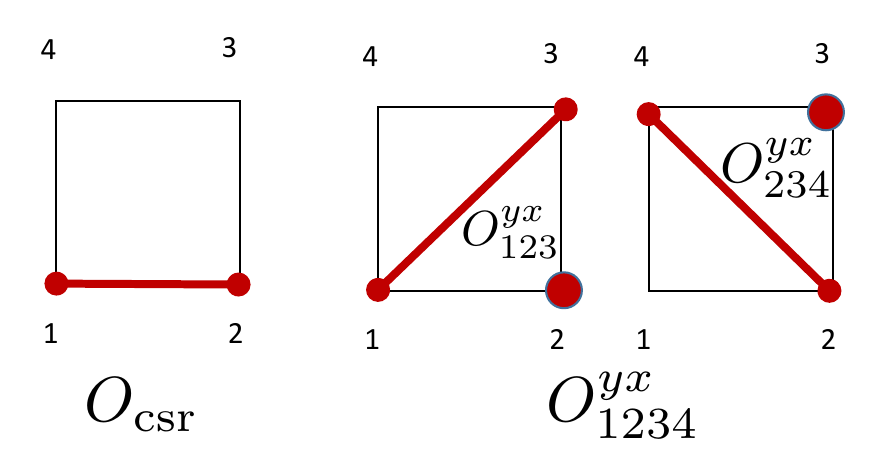}
\caption{Operators which are averaged over in the CSR and CMC. Line denotes a bond operator $(S^{+}_{i}S^{-}_{j}+(i \leftrightarrow j))$ 
and the circle  denotes a site operator $S^{z}_{i}$, respectively.  }
\label{fig:Operators}
\end{center}
\end{figure}

The coefficients in the high temperature expansion of the susceptibilities are found using a numerical symbolic multiplication method, described in Appendix \ref{App:Auto}. Here, we outline the scheme by two simplest examples- the CSR and CMC, defined in Eq.~(\ref{CSR-c}) and Eq.~(\ref{CMC-c}) respectively.   

\subsection{CSR}
\label{App:CSR}
\begin{figure}[h]
\begin{center}
\includegraphics[width=8cm,angle=0]{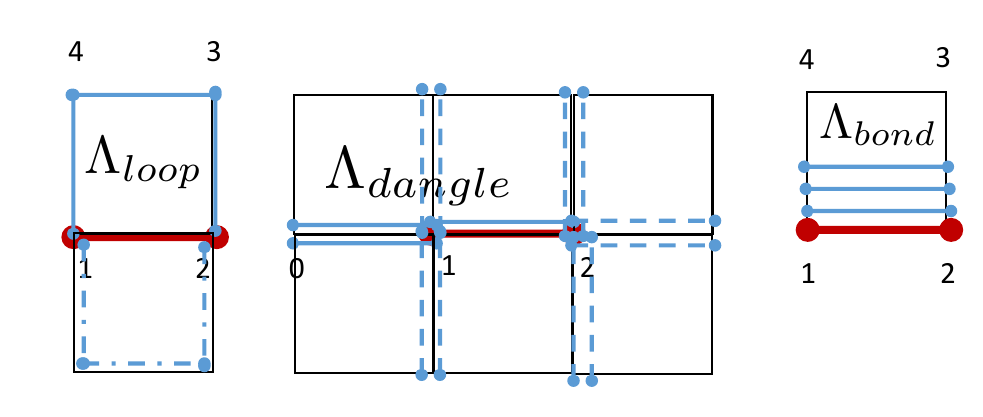}
\caption{Graphs of operators which contribute to CSR at $\cO(\beta^3)$. }
\label{fig:CSR-3}
\end{center}
\end{figure}

The CSR is given by averaging the single bond operator, shown in Fig.~\ref{fig:Operators}.
The first two leading orders in $(\beta t)$ of the CSR are defined as,
\be
\bar{ \chi}_{\rm csr} = q^2 \beta \tt^2 \bar{\chi}^{(1)}_{\rm csr} + q^2 \beta^3\tt^4 \bar{\chi}^{(3)}_{\rm csr} +\ldots.
\label{chi-CSR}
\ee
where, 
\be
\bar{\chi}_{\rm csr}^{(1)} = \frac{1}{2}(1-m^2) .
\label{chi-CSR1}
\ee
The graphs contributing to the order $\beta^3$ CSR are shown in Fig.~\ref{fig:CSR-3}. The result (using symbolic multiplication) is- 
\be 
\bar{\chi}_{\rm csr}^{(3)} (\delta) = \frac{\beta^3}{24}(-1+20 \delta^2 -19\delta^4).
\label{chi-CSR3}
\ee 
Upon transforming this using Eq.~(\ref{corr}), one gets-
\bea
\chi_{\rm csr}^{(3)} (m)&=& \lim_{\delta\to m}\{  \chi_{\rm csr}^{(3)} (\delta) -(\partial_\delta \chi^{(1)}_{\rm csr}) \delta (1 - \delta^2)\} \nonumber\\
&=& {1 \over 24}  (-1 -4m^2 +5m^4).\nonumber\\
\eea

Finally, if one substitutes $m=2(n-{1\over2})$ in Eq.~\ref{chi-CSR1} and Eq.~\ref{chi-CSR3} in Eq.~\ref{chi-CSR}, we obtain Eq.~(\ref{CSR-c}).

\subsection{CMC}
The CMC is given by averaging the plaquette operator, shown in Fig.~\ref{fig:Operators}.
\bea
\chi_{\rm cmc} &=& - 2 q^3  t^2 \langle O^{yx}_{123}\rangle \nonumber\\
&=&  2 q^3  t^2 \langle  \left(  (S^{+}_{1}S^{-}_{3}+S^{-}_{1} S^{+}_{3})S^{z}_{2} \right)\rangle , 
\label{chi-CMC}
\eea
where we have used using C4 symmetry to equate  four identical contributions to the  expectation value. 
The leading order $\chi^{(2)}_{\rm cmc}$ requires tracing $ O^{yx}_{123}$ times two Hamiltonian bonds in a connected cluster inside a plaquette.
The calculation yields 
\bea
\chi^{(2)}_{\rm cmc} &=&  -  \Tr \rho_0  (\beta H)^2 O^{yx}_{123} \nonumber\\
&=& -{(\beta\tt)^{2} \over 2}m(1-m^2).
\label{chi-CMC2} 
\eea

\begin{figure}[h]
\begin{center}
\includegraphics[width=6cm,angle=0]{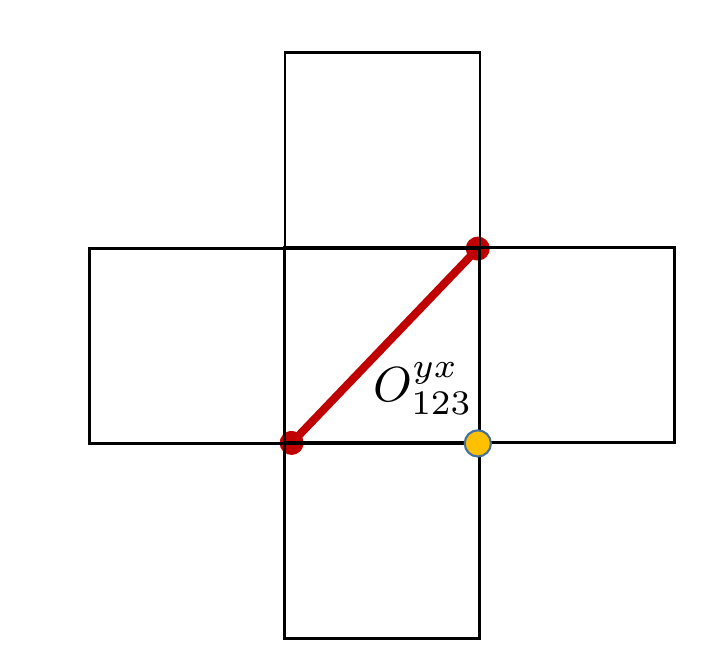}
\caption{Minimal connected cluster for the   calculation of   $\chi^{(4)}_{\rm cmc}$, which must include 4 powers of $H$ whose bonds connect to the sites of $O^{yx}_{123}$
and contribute to  a non vanishing trace.}
\label{fig:CMC-4}
\end{center}
\end{figure}
The order $\beta^4$ contribution to CMC is obtained to be- 
\be
\chi_{\rm cmc}^{(4)} (\delta) =   {(\beta t)^4 \over 24} (15\delta - 42\delta^3 + 27\delta^5) .
\label{chi-CMC4}
\ee
The minimal cluster chosen for this computation is shown in Fig.~\ref{fig:CMC-4}. 
Transforming the expression using Eq.~(\ref{corr}), yields
\bea
\chi_{\rm cmc}^{(4)} (m)&=& \lim_{\delta\to m}\{  \chi_{\rm cmc}^{(4)} (\delta) -(\partial_\delta \chi^{(2)}_{\rm cmc}) \delta (1 - \delta^2)\} \nonumber\\
&=& {(\beta t)^4 m \over 8}  (1 - m^2) (1 + 3m^2).\nonumber\\
\eea 

Again, combining Eq.~\ref{chi-CMC2} and Eq.~\ref{chi-CMC4}, replacing $m=2(n-{1 \over 2})$  in Eq.~\ref{chi-CMC} we obtain Eq.~(\ref{CMC-c}).

\subsection{Thermal susceptibilities}
\label{highTexpth}
The thermal currents are defined in Eq.~(\ref{jQ}). In the CMC susceptibilities used in Eq.~(\ref{chi-Q}) we use  the electric magnetization $M$ as defined in Eq.~(\ref{electric}). The chemical potential $\mu$ at high temperatures is related to the density by solving the single-site problem (neglecting the $\beta H$ term), which yields Eq.~(\ref{muT}),
\be
n ={e^{\beta\mu} \over 1+e^{\beta\mu}}\Rightarrow \mu(\beta) =\beta^{-1} \log\left({n\over 1-n}\right).
\ee

The new susceptibilities required to determine the thermal Hall coefficient 
($\chi_{\rm csr}^{\rm ee}$,$\chi_{\rm cmc}^{\rm ee}$,$\chi_{\rm csr}^{\rm ec}$ and $\chi_{\rm cmc}^{\rm ec}$)  were all evaluated using the methods explained in Appendix~\ref{App:Auto}.
The energy current itself turns out to be a combination of three-site operators, containing next-nearest neighbour currents decorated with $S^z$ operators at nearest-neighbour locations. Hence, the resulting susceptibilities turn out to be more complicated and have different $\beta$ dependencies to leading order compared to the charge case. One finds the following expressions- 
\bea
\chi^{\rm ee}_{\rm csr} &=& \beta t^4 \left( 1 - {m^2 \over 2} - {m^4 \over 2} \right), \nonumber \\
\chi^{\rm ee}_{\rm cmc} &=& {q\beta^2 t^6 \over c} \left( {-7m \over 2} + m^3 + {5m^5 \over 2} \right), \nonumber\\ 
\chi^{\rm ec}_{\rm csr} &=& {3q\beta t^4\over 2} m\left( 1 - m^2\right), \nonumber \\
\chi^{\rm ec}_{\rm cmc} &=& - {q^2\beta t^4\over 2} \left( 1 - m^4 \right).
\eea
Now, at low $\beta$ (high temperatures), Eq.~(\ref{chi-Q}) shows that $\chi_{\rm csr}^Q$ is dominated by $({\mu \over q})^2\chi_{\rm csr}$ (the charge CSR term), which is $O(1/\beta)$ and both $-2({\mu \over q})\chi^{\rm ec}_{\rm cmc}$ and $({\mu \over q})^2\chi_{\rm cmc}$ contributions are important ($O(1)$) in $\chi_{\rm cmc}^Q$. 

However, one also observes that the susceptibilities (as well as $\mu$) have different density dependencies near half-filling.
For instance, $\mu$, $\chi_{\rm csr}^{\rm ec}$, $\chi_{\rm cmc}^{\rm ee}$ and $\chi_{\rm cmc}$ cross zero at $n={1 \over 2}$ by particle-hole symmetry / anti-symmetry, while the thermoelectric CMC ($\chi_{\rm cmc}^{ec}$) is a very flat function of $n$ near half-filling. Hence, the sign of the thermal Hall coefficient is determined by $\chi_{\rm ec}^{\rm cmc}$ and turns out to be opposite to the charge case near $n={1 \over 2}$. 

\section{Computing the correction term}
\label{App:Corr}

\begin{table}[t]
\begin{tabular}{c c p{8cm}}
\hline
$a$ & $b$ & $\chi_{\rm csr} \tilde{M}_{ab}/(\beta^2 q^3 t^{(a+b+4)})$ \\
\hline
0 & 0 & $-m(1-m^{2})$ \\
0 & 2 & $-\frac{1}{2}m(1-m^{2})^{2}$ \\
0 & 4 & $-\frac{1}{2}(56m+181m^{3}-194m^{5}+69m^{7})$ \\
2 & 0 & $-m(1-m^{2})^{2}$ \\
2 & 2 & $\frac{1}{2}(36m-53m^{3}-2m^5+19m^7)$ \\
2 & 4 & $\frac{1}{2}(1871m-2638m^{3}-1017m^{5}+2464m^{7}-680m^{9})$ \\
4 & 0 & $m(1-m^{2})^{3}$ \\
4 & 2 & $\frac{1}{2}(1307{m}-2476m^{3}+969m^{5}+262m^{7}-62m^{9})$\\
4 & 4 & $\frac{1}{2}(115281m-240247m^{3}+109284m^{5}+39594m^{7}-22457m^{9}-1455m^{11})$ \\
\hline
\end{tabular}
\caption{Charge hypermagnetization matrix elements (unnormalized) at $T=\infty$.}
\label{table:Mcab0}
\end{table}

The normalized matrix elements of Eq.~(\ref{Rcorr}) require expanding the  operators  $\cL^{2k} j^\alpha $ in the
Krylov basis $|k;\alpha\rangle$~\cite{EMT}. Using
\be
\cL^{k} j^\alpha = \sum_{k'} |k';\alpha\rangle\langle k';\alpha| \cL^{k}  |0;\alpha\rangle= \chi_{\rm csr} \sum_{k'} K_{k,k'} |k';\alpha\rangle
\ee
where $K_{k',k}= (L^k)_{k',0}$ are functions of a finite number of recurrents, $\Delta_1,\ldots \Delta_{k'\le k}$.
Using
\be
|k,\alpha\rangle = K^{-1}_{k,k'} \cL^{k'} |0\rangle
\ee
we can transform any normalized matrix element of the hypermagnetization into unnormalized matrix elements
\be
\tilde{M}_{n'm'} = (0,y|\cL^{n'} \cM \cL^{m'}|0,x),
\ee
which are easier to calculate, using
\be
M_{n,m}'' = \sum_{n'\le n}\sum_{m'\le m} K^{-1}_{n,n'}K^{-1}_{m,m'}  \tilde{M}''_{n',m'}. 
\ee
The latter are computed using the methods described in Appendix~\ref{App:Auto} are quoted in Table~\ref{table:Mcab0}.

\section{Automated evaluation of operator traces}
\label{App:Auto}

High order moments $\mu_{2k},k=0\ldots 5$,
recurrents $\Delta_{k<5}$ and hypermagnetization matrix elements $\tilde{M}_{nm},nm\le 4$ require traces over a large number of site-operators products on square lattice.
We've used symbolic multiplication to perform these traces. 

The clusters are formed by commuting bond operators of the Hamiltonian or magnetization with the root current operator $j_{\langle ij\rangle}^\alpha$ on a single bond $\langle ij\rangle$ (utilizing translational symmetry). The result of $\cL^n j^\alpha_i$ is a sum of multi-site products of operators $O_{i_1}({\br_1}) \cdot  O_{i_2}(\br_2)\cdot ... \cdot O_{i_N}(\br_N)$, which is treated as a new ``hyperstate''  with a complex amplitude that is stored separately.

The rapid (factorial) growth of the number of such operator products limited extending  our calculations beyond $n=4$.
At fourth order we calculated traces over $\sim 10^{7}$ operators.

The new hyperstates are generated using  
the site-local operator multiplication table,
\be
\sigma_{i}\sigma_{j}=\delta_{ij}+i\epsilon_{ijk}\sigma_{k}, 
\ee
where $\sigma_{i}$'s are usual Pauli spin matrices. We performed the calculations on a $24\times24$ square grid, which ensured that the even the largest operator products were contained within the lattice. After constructing the operators, we multiplied them with powers of the Hamiltonian (also a collection of bond operators) to ensure that they had a non-vanishing trace. Most of the resuting hyperstates have zero trace, owing to unmatched Pauli matrices
\be
Tr(\sigma^x_{i})=Tr(\sigma^{y}_{i})=0.
\ee
The computation was reduced by eliminating operators with unmatched $\sigma^x$'s or $\sigma^y$'s. Moreover, we  combined equal operator products at intermediate stages, to  retain only distinct hyperstates. This sorting becomes time consuming at about $~10^{6}$ operators.
Finally, we evaluated the traces over the remaining density ($\sigma^z$) factors,
\be
Tr(\sigma^z_{i})=\delta=m|_{\beta=0},
\ee
as given in Appendix~\ref{App:HiT}.

\section{QMC calculation of the Hall coefficient}
\label{QMC_RH}

In this short section, we provide a few details of the QMC calculation, used to generate the lower temperature results for $R_{\rm H}^{(0)}$, as depicted in Fig.~\ref{fig:RHvsT}.
We've used the DSQSS package~\cite{Motoyama} for this purpose. This package employs a path-integral based Monte Carlo scheme for bosons and quantum spin systems, devoid of a sign problem. 
More specifically, it uses the Directed Loop Algorithm (DLA)~\cite{DLA}. 
In DLA, one adds a pair of `worm heads' at two randomly chosen space-time points within the simulation. 
These represent insertions of off-diagonal operators (like $S^{x}$). Thereafter, one of these is kept fixed (called the `tail'), 
while the other `propagates' obeying detailed balance rules. In course of propagation, the head scatters off `vertices', which are two-spin operators specific to the Hamiltonian in question. 
Finally, one terminates the propagation when the head meets the tail. The `closed loop' configurations thus obtained contribute to the free energy. The statistics obtained from the worm propagation moves (while the loop is open) corresponding to a fixed space-time separation of the worm head and tail directly provide an estimate for the two-point off-diagonal correlation functions
(like $<S^+_{i}(\tau)S^-_{j}(0)>$). We used the equal-time ($\tau=0$) results for our purpose. For the operators shown in Fig.~\ref{fig:Operators}, one can directly use this result (for nearest-neighbour separation) to calculate the CSR, while for the CMC, one has to
`re-weight' the measurements (for next-nearest neighbour separation) by the $S^{z}$ value of the spin at the nearest neighbour location. We achieved this by slightly modifying the correlation function measurement part of the source code.
\bibliographystyle{unsrt}
\bibliography{refs.bib}

\begin{thebibliography}{10}
\bibitem{Comm:Homes}
{This behavior is also consistent with a combination of Uemura's scaling of
  transition temperature with superfluid stiffness~\cite{Uemura}, , the linear
  resistivity of clean optimally doped samples above $T_{\rm c}$, and Homes law
  relation between superfluid stiffness and resistivity $R_{xx}(T_{\rm
  c})$~\cite{Homes} }.
\end{thebibliography}
\end{document}